\documentclass[12pt]{iopart}

\pdfoutput=1

\usepackage{iopams}
\usepackage{setstack}
\usepackage{parskip}
\usepackage[]{graphicx}

\usepackage{url}

\usepackage{geometry}
\usepackage{fancyhdr}
\begin{document}

\title[Visual guide to optical tweezers]{Visual guide to optical tweezers}
\author{Isaac C D Lenton, Alexander B Stilgoe, \\Halina Rubinsztein-Dunlop,
    and Timo A Nieminen}
\address{The University of Queensland,
    School of Mathematics and Physics, Brisbane, QLD 4072, Australia}
\ead{timo@physics.uq.edu.au (T A Nieminen)\\
     $\qquad\ $ stilgoe@physics.uq.edu.au (A B Stilgoe)}

\begin{abstract}
It is common to introduce optical tweezers using either geometric optics
for large particles or the Rayleigh approximation for very small
particles.
These approaches are successful at conveying the key ideas
behind optical tweezers in their respective regimes.
However, they are insufficient for modelling particles of
intermediate size and large particles with small features.
For this, a full field approach provides greater insight into
the mechanisms involved in trapping.
The advances in computational capability over the last decade
has led to better modelling and understanding of optical tweezers.
Problems that were previously
difficult to model computationally can now be solved using a
variety of methods on modern systems.
These advances in computational power allow for full field solutions
to be visualised, leading to increased understanding of the
fields and behaviour in various scenarios.
In this paper we describe the operation of optical tweezers using
full field simulations calculated using the finite difference
time domain method.
We use these simulations to visually illustrate various
situations relevant to optical tweezers, from the basic operation
of optical tweezers, to engineered particles and evanescent fields.
\end{abstract}


\textbf{Preprint of:}\\
Isaac C. D. Lenton, Alexander B. Stilgoe, Halina Rubinsztein-Dunlop and Timo A. Nieminen\\
``Visual guide to optical tweezers''\\
\textit{European Journal of Physics} \textbf{38}(3), 034009 (2017)\\
\url{https://doi.org/10.1088/1361-6404/aa6271}

\maketitle

\section*{~}

\section{Introduction}

Development of optical tweezers began in the second half of the
20th centry with the first demonstrations of optical levitation
in 1971 and the single beam gradient trap in 1986 \cite{ashkin71, ashkin86}.
It had long been known that light could influence the trajectories of
particles, proposals for solar sails had been recorded
as early as 1924 \cite{tsander24}, but the small momentum carried by
photons ($\hbar\omega/c$ per photon, compared to $\hbar\omega$ energy, where
$\omega$ is the optical angular frequency and $c$ is the speed of light
in free space)
made experimental demonstrations of
optical forces difficult without high intensity light sources.
The invention of the laser in 1960 likely contributed to
the realisation of optical levitation and optical tweezing by
allowing high intensity spatially confined beams to be used in experiments.
Since their invention, optical tweezers have seen applications to
various fields ranging from microbiology where they are used to
trap and manipulate cells and DNA to microfluidics where diffractive
optical elements can be combined with optical tweezers to measure
properties of fluids such as viscosity and elasticity
\cite{wang97, yao09, nieminen05}.

The concepts behind the operation of optical tweezers are often
introduced using either geometric optics suitable for large spherical
particles \cite{ashkin86}
or the Rayleigh approximation for small dipole-like particles \cite{ashkin86}.
Both these methods can provide important insight into the basic
principles behind optical tweezers. In particular, both give a picture
where the optical force can be separated into a gradient force and a
scattering force.
The Rayleigh approximation assumes particles are much smaller
than the illumination wavelength.
In this regime the particle are modelled as a single scattering dipole.
for which there are analytical solutions.
The geometric optics approach works well for large particles where
interference effects and minimum beam radius can be ignored.
Reference \cite{callegari15} has some very nice animations and
figures generated using geometric optics; their toolkit is
suitable for teaching OT.
However, for intermediate particles, large particles with small features
or trapping involving evanescent fields, the full electromagnetic
fields give much greater insight into the operation of optical tweezers.

Apart from a few specific cases for certain regimes, there are no
analytical models for optical tweezers or the fields in/around
arbitrary particles.
As such, numerical modelling
is an important tool for the design and analysis of tweezers experiments
\cite{nieminen14}, good agreement between experimental
and computational results has been demonstrated \cite{fontes05, rohrbach05}.
Computational modelling of optical tweezers involves calculating how
light is scattered by a trapped object in order to calculate
forces, torques and other properties of interest.
The advances in computational power and availability of codes and
algorithms for modelling optical tweezers in the recent decade
has led to increased ability to model optical tweezers.
Many methods exist for performing scattering calculations \cite{wriedt98}.
However, many of the codes implementing these methods are not directly
applicable to optical tweezers problems because they either don't
calculate properties of interest to optical tweezers
or they make assumptions such as plane wave illumination \cite{nieminen04}.

The advances in computational capability and the availability of
suitable codes for simulating optical tweezers has led to
better models of optical tweezers.
Some of these methods had been previously developed but suitable
codes for modelling optical tweezers situations with suitable
illumination have not always been available.
Every numerical method has a unique set of advantages and limitations;
a brief review of methods used by our group can be
found in \cite{bui2016}.
Our group currently uses a range
of methods to simulate optical tweezers including the discrete dipole
approximation, finite difference frequency domain and various Generalized
Lorenz--Mie Theory and point matching methods.
We have recently developed a finite difference time domain (FDTD)
implementation, which solves Maxwell's equations in the time domain
by calculating discrete differences of the fields.
FDTD is convenient for visualising the fields since the full fields
are calculated as a convenient consequence of the method.

For this work, FDTD was chosen as this method gives the real
fields and can easily simulate arbitrary geometries and
inhomogeneous dielectric and conductive materials.
Commercial FDTD tools are available that can be configured to
simulate optical tweezers but our group has been developing
our own FDTD package which we hope to release as open source software.
The benefits of having an open source package are the ability to
modify and configure the code in order to allow testing of new
methods and calculation of properties of interest.
The core components to the FDTD implementation are fairly straightforward;
2-D or 3-D implementations used as a learning tool or the development
of a 2-D implementation as a guided exercise could make interesting
educational tools.

The remainder of this paper is split into three parts.
\Sref{sec:ot} provides an overview of optical tweezers
including explanations of their operation using
the Rayleigh approximation, geometric optics approach and
full field simulations.
\Sref{sec:fdtd} briefly describes FDTD and the
types of illumination commonly used for optical tweezers.
Finally, \sref{sec:simulations} presents various simulations
generated using FDTD which illustrate different optical
trapping scenarios.

\section{Optical tweezers}
\label{sec:ot}

\begin{figure}[ht]
\centering
\includegraphics[width=\textwidth]{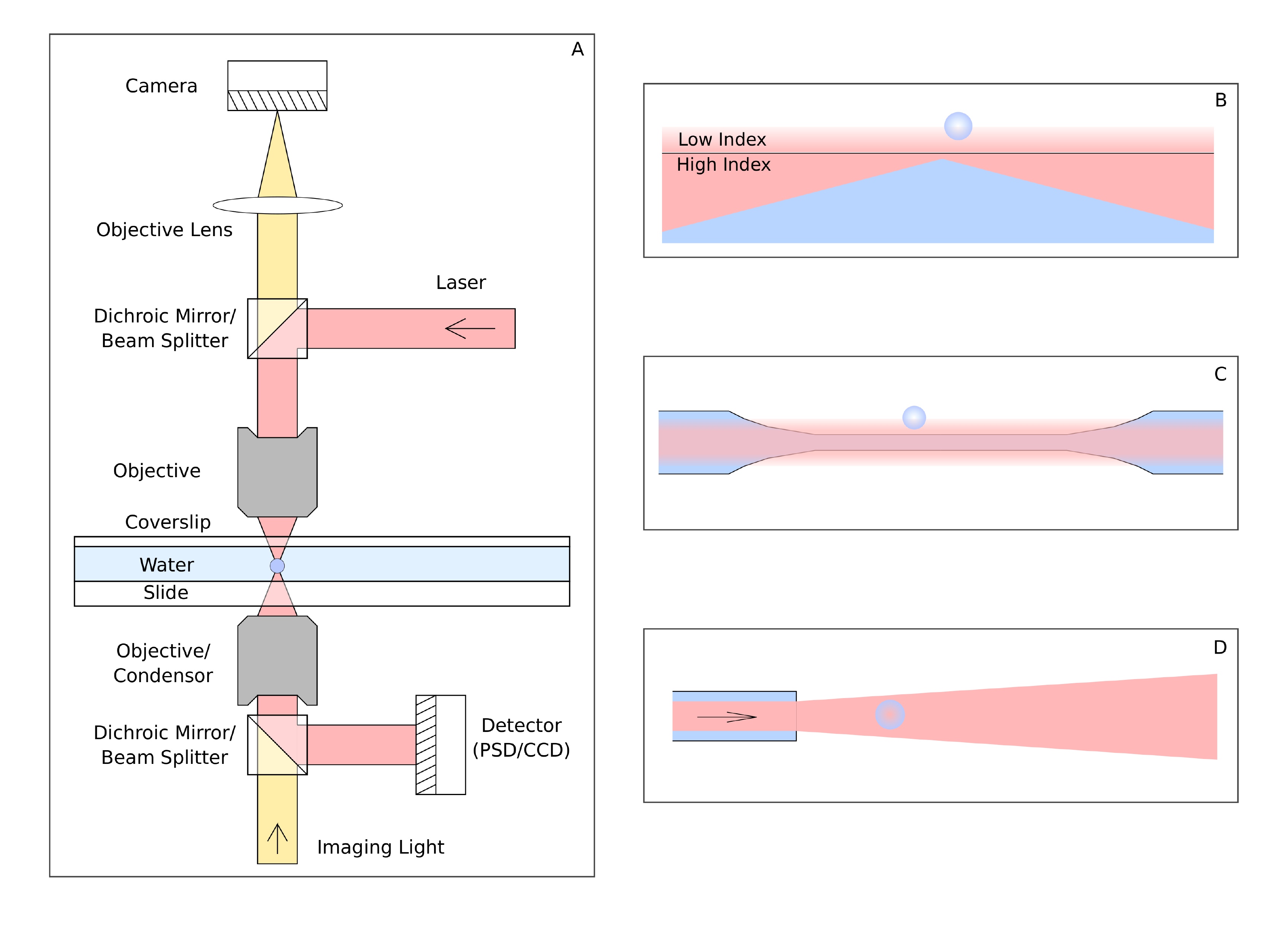}
\caption{(A), a typical optical tweezers apparatus.
    (B), a particle
    in an evanescent field near the interface of a high/low index material.
    (C) a particle in an evanescent field from a tapered optical fibre.
    (D) a fibre based optical tweezers system.}
\label{fig:apparatus}
\end{figure}

An optical tweezers experiment requires: a light source, such as a
tightly focussed laser beam; a system to be investigated, such as a
particle suspended in water; and a detection system to measure the
properties of interest.
A typical optical tweezers apparatus is shown in \fref{fig:apparatus}-A.
This apparatus
consists of a microscope system with the optical trapping beam
introduced through the microscope objective.
The particle to be trapped is located on a microscope slide
between the microscope objective and the condenser.
This kind of system allows the position of the particle to be directly
imaged or the position to be inferred from the deflection of the
trapping beam after passing through the system.
Detection systems include quadrant photodiodes, high resolution
cameras and position sensitive detectors.
The system can be extended to include multiple optical traps by
using multiple light sources, by time-sharing
a beam over multiple locations with an acousto-optic modulator
or by splitting the beam with diffractive or reflective optical elements
such as a digital micro-mirror device or spatial light modulator
\cite{biochemistryReview}.
In addition to the system shown in \fref{fig:apparatus}-A,
other optical trapping systems may involve the use of evanescent
fields or optical fibres or a combination of multiple systems
depending on the problem requirements.

There are multiple forces present in a typical optical tweezers system;
in addition to the optical force from the trapping beam other forces
from gravity, buoyancy, Brownian motion or other properties of the fluid
might also apply \cite{bui2016}.


\subsection{Full field simulation and visualisation}

There are two convenient starting points for
calculation of optical forces in a full field simulation:
an electromagnetic force law (e.g., the Lorentz force law),
and the momentum flux density (the Maxwell stress tensor). We will
begin with the first of these. The force density $\mathbf{f}$ exerted
by electromagnetic fields on matter is, according the Lorentz force law:
\begin{equation}
\mathbf{f} = \rho \mathbf{E} + \mathbf{J} \times  \mathbf{B},
\end{equation}
where the force density is given by the forces acting on the
charge density, $\rho$, and the current density, $\mathbf{J}$. For a dielectric
particle such as is typically trapped in optical tweezers, we can
replace the dielectric polarisation $\mathbf{P}$ per unit volume by its
equivalent charge density $\rho = -\nabla\cdot\mathbf{P}$, and the polarisation
current $\mathrm{d}\mathbf{P}/\mathrm{d}t$, the Lorentz force law becomes
\begin{equation}
\mathbf{f} = (\mathbf{P}\cdot\nabla)\mathbf{E}
+ \frac{\mathrm{d}\mathbf{P}}{\mathrm{d}t}\times\mathbf{B}.
\end{equation}
We can then use the identity
$(\mathbf{E}\cdot\nabla)\mathbf{E} = \nabla(\mathbf{E}\cdot\mathbf{E})/2
- \mathbf{E}\times(\mathbf{E}\times\mathbf{E})$, the relationship between $\mathbf{P}$ and 
$\mathbf{E}$, $\mathbf{P} = (\epsilon - \epsilon_0 )\mathbf{E}$, where
$\epsilon$ and $\epsilon_0$ are the permittivity of the particle and of
free space, respectively, and the Maxwell equations \cite{gordon1973} to obtain
\begin{equation}
\mathbf{f} = (\epsilon - \epsilon_0) \left( \frac{1}{2} \nabla(\mathbf{E}\cdot\mathbf{E})
 + \frac{\mathrm{d}}{\mathrm{d}t}( \mathbf{E}\times\mathbf{B}) \right).
\end{equation}
A similar force also acts on the surrounding medium, if it has a permittivity different
from $\epsilon_0$, and we must account for the force from contact with the
surrounding medium. This gives an effective force density is
\begin{equation}
\mathbf{f} = (\epsilon_\mathrm{particle} - \epsilon_\mathrm{medium}) \left( \frac{1}{2} \nabla(\mathbf{E}\cdot\mathbf{E})
 + \frac{\mathrm{d}}{\mathrm{d}t}( \mathbf{E}\times\mathbf{B}) \right).
\label{eq:force-density}
\end{equation}

\subsubsection{Rayleigh approximation}

For the case where a particle is much smaller than the wavelength
(a Rayleigh particle), the applied field
can be assumed to be uniform over the particle, allowing us to very easily integrate
the force density over the particle to find the total force:
\begin{equation}
\mathbf{F} = \mathbf{f} V,
\end{equation}
where $V$ is the volume of the particle. For a sphere of radius $a$,
$V = (4/3)\pi a^3$, and we can express the total force in terms of a polarisability $\alpha$:
\begin{equation}
\mathbf{F} = \alpha \left( \frac{1}{2} \nabla(\mathbf{E}\cdot\mathbf{E})
 + \frac{\mathrm{d}}{\mathrm{d}t}( \mathbf{E}\times\mathbf{B}) \right).
\label{force_dipole}
\end{equation}
The static polarisability of a sphere,
\begin{equation}
\alpha_\mathbf{static} = 4\pi\epsilon_\mathrm{medium} a^3
\left( \frac{\epsilon_\mathrm{particle}/\epsilon_\mathrm{medium}-1}
{2\epsilon_\mathrm{particle}/\epsilon_\mathrm{medium} + 1} \right),
\end{equation}
is a good starting approximation
for the polarisability $\alpha$.
For a time-harmonic incident field, the time-averaged force is
\begin{equation}
\langle\mathbf{F}\rangle = \frac{1}{2} \mathrm{Re} \left\{ \alpha \left( \frac{1}{2} \nabla(\mathbf{E}_0\cdot\mathbf{E}_0^\ast)
 - 2 \mathrm{i}\omega \mathbf{E}_0\times\mathbf{B}_0^\ast \right) \right\}.
\label{force_dipole_average}
\end{equation}
The first term, proportional to the gradient of the irradiance, is the gradient force,
and the second term, proportional to the Poynting vector, is the scattering force.
Using the static polarisability leads to a time-averaged scattering force
of zero (because the polarisability is real). However, a time-harmonic dipole
moment of amplitude $\mathbf{p}_0$ radiates an average
power of $c^2 Z k^4 \mathbf{p}_0\cdot\mathbf{p}_0^\ast / (12\pi)$, where $k = 2\pi/\lambda$ is the
wavenumber and $Z$ the impedance of the medium, the polarisability must
have an imaginary component. From the radiated power, the imaginary component of the polarisability must be
\begin{equation}
\mathrm{Im}(\alpha) = -\frac{cZk^3}{6\pi}|\alpha|^2,
\end{equation}
giving the required scattering force.
For non-spherical Rayleigh particles, the appropriate static polarisability tensor can be
used, correcting for radiated power.

Giving a simple analytical expression for the force on a trapped particle, depending
only on the local values of the electromagnetic field and derivatives, the Rayleigh
approximation can provide much useful insight into the optical trapping of small
particles, such as the relative scaling of the gradient force, scattering force,
and Brownian motion \cite{ashkin86}.
However, it is only quantitatively accurate for
very small particles. In principle, we could integrate the
effective force density given by \eref{eq:force-density}
over the volume of the particle.  We can also approach the
problem differently, in terms of the momentum flux density, as given by the
Maxwell stress tensor (MST).

\subsubsection{Maxwell stress tensor}

A general approach to calculating the optical force on a region
involves integration of the Maxwell stress tensor and
Poynting vector over a surface/volume surrounding the particle
\cite{nieminen14}
\begin{equation}
\mathbf{F} = \int_S \bar{\mathbf{T}}\cdot \mathrm{d}\mathbf{S}
		- \epsilon\mu\frac{\mathrm{d}}{\mathrm{d}t}
    \int_V \mathbf{U} \mathrm{d}V
\label{eq:general-force}
\end{equation}
where $S$ is a surface surrounding the volume, $V$, containing
the particle; $\mathbf{U}$ is the Poynting vector and $\bar{\mathbf{T}}$
is the MST which can be defined using
the Kronecker delta function $\delta_{ij}$ as
\begin{equation}
\bar{\mathbf{T}}_{ij} = \epsilon \mathbf{E}_i\mathbf{E}_j + \mu \mathbf{H}_i \mathbf{H}_j
    - \frac{1}{2}\left( \epsilon |\mathbf{E}|^2 + \mu |\mathbf{H}|^2 \right)\delta_{ij}.
\label{eq:mst}
\end{equation}

The above equation gives the sum of the electromagnetic force
acting on object contained in the volume and the rate of change
of electromagnetic momentum within the volume. For a time-harmonic
field, we can eliminate this second part by taking the time
average, then \eref{eq:general-force} reduces to
\begin{equation}
\langle \mathbf{F} \rangle = \int_S \langle\bar{\mathbf{T}}\rangle\cdot\mathrm{d}\mathbf{S}.
\label{eq:time-averaged-force}
\end{equation}
Almost always, we wish to know the time-averaged force rather than the
instantaneous electromagnetic force, so there is no practical loss of
generality.
The force can also be calculated using a similar volume integral form
\begin{equation}
\langle \mathbf{F} \rangle = \int_V \nabla\cdot\langle\bar{\mathbf{T}}\rangle\mathrm{d}V
\label{eq:volume-integral-force}
\end{equation}
where $\nabla\cdot\langle\bar{\mathbf{T}}\rangle$ denotes the divergence of the MST.
For a particle in continuous wave illumination, this formula gives
the average force when the time average is taken over one complete
optical cycle.

For pulsed illumination, such as when using a femtosecond laser, the
time average needs to be taken over a complete pulse.
To calculate the average optical force on the particle the MST
integral needs to be evaluated at multiple time steps.
For a continuous wave source, the time average need only be taken
over multiple samples from a single optical cycle.
The error in the time average typically scales better than $1/m$
where $m$ is the number of samples.
We have found that for most cases $m = 3$ produces satisfactory results.

\subsubsection{Geometric optics}

We have not specified the surface of integration for
\eref{eq:time-averaged-force}. In principle, any surface enclosing
the particle of interest, and containing nothing else experiencing
a net force from the beam, can be used. A particular simple choice
for a focussed beam. For such a beam, a well-defined far field exists,
where the beam becomes a spherical wave. Taking the far field limit,
it is convenient to use a spherical coordinate system $(r,\theta,\phi)$
for the calculation,
since our surface area element $\mathrm{d}\mathbf{S} = r^2\mathrm{d}\theta\mathrm{d}\phi$,
and radial components of the electromagnetic field can be neglected, and our integral
reduces to
\begin{equation}
\mathbf{F} = - \int_S\frac{1}{2}\left( \epsilon |\mathbf{E}|^2 + \mu |\mathbf{H}|^2 \right) r^2\mathrm{d}\theta\mathrm{d}\phi.
\end{equation}
The integrand is the energy density; in the far field, the energy density $u$,
the energy
flux density (the Poynting vector $\mathbf{U}$), the momentum density $\mathbf{g}$, and the radial momentum flux density
$T_{rr}$ are very simply related:
\begin{eqnarray}
u & = & |\mathbf{U}|/c \\
|\mathbf{g}| & = & u/c = |\mathbf{U}|/c^2 \\
T_{rr} & = & u.
\end{eqnarray}
This is an old result, first given by Maxwell (with a factor of $1/3$, for isotropic rather than
collimated radiation). Thus, we could also choose to integrate the Poynting vector rather than the energy density.

\begin{figure}[ht]
\centering
\includegraphics[width=\textwidth]{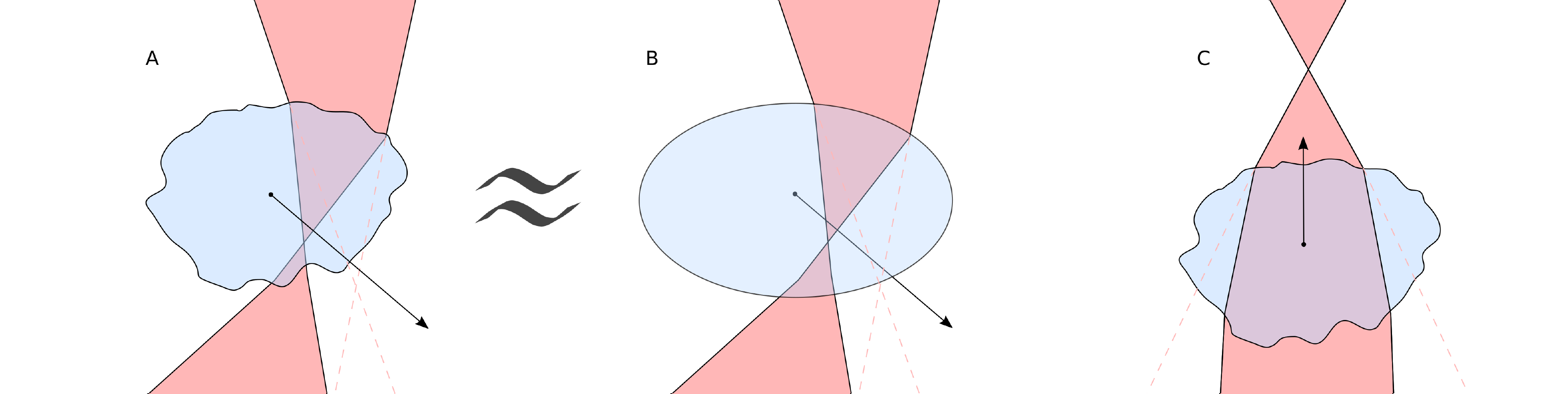}
\caption{Ray diagrams showing deflection of a beam by a particle.
    The refractive index of the particle
    is higher than the refractive index of the background medium.
    (A) shows a particle with a rough surface, if the
    surface deviation is small compared to the particle size the
    situation in (A) is approximately equivalent to
    a similarly sized spherical particle (B).
    (C) shows a particle that makes the beam more collimated,
    the corresponding force experienced by the particle is opposite the beam
    propagation direction.}
\label{fig:ashkin-ish-diagrams}
\end{figure}

Apart from the far field giving simple expressions for the transport of energy and momentum, it also allows
us to represent the field in terms of rays---the geometric optics approximation. In this case, we can use ray
tracing to calculate the interaction between the field and the particle.
This comes with some costs:
in the geometric optics approximation, we neglect diffraction and interference effects, and the
particle must be large compared to the wavelength (and features on the particle must also be large compared
to the wavelength). Potentially worse, surfaces of the particle must not be near the focus of
the beam, since the focal region is not accurately described in the approximation. Nonetheless,
geometric optics can still given a useful qualitative view of optical trapping, as shown
in \fref{fig:ashkin-ish-diagrams}, and often useful quantitative results as well.

\subsubsection{Between the Rayleigh and geometric optics regimes}

\begin{figure}[ht]
\centering
\includegraphics[width=0.24\textwidth]{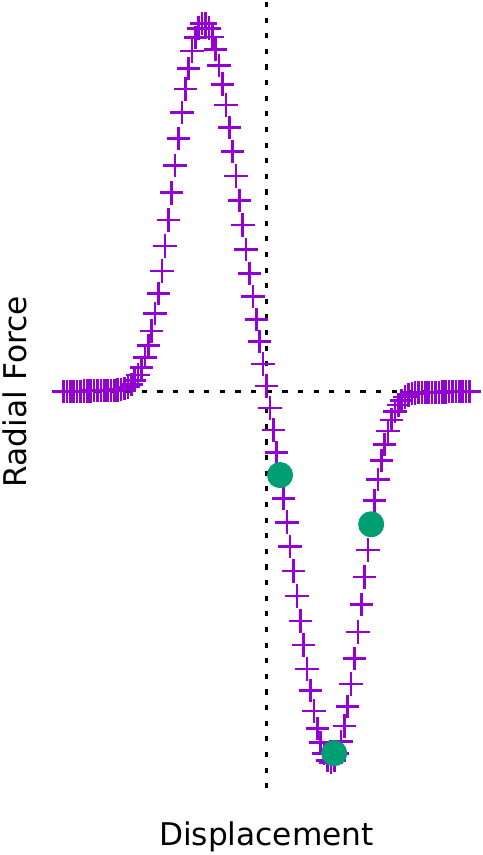}\hfill
\includegraphics[width=0.24\textwidth]{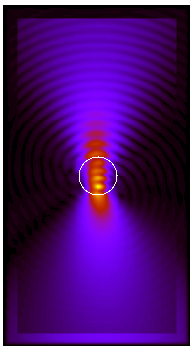}\hfill
\includegraphics[width=0.24\textwidth]{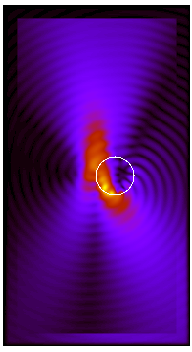}\hfill
\includegraphics[width=0.24\textwidth]{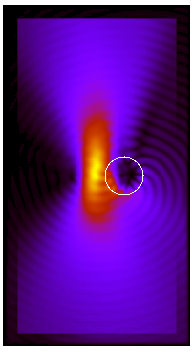}
\caption{Deflection of a weakly focussed beam by a dielectric particle.
    The images on the right correspond to the three locations
    marked on the force-displacement graph.}
\label{fig:deflection}
\end{figure}

Many of the particles of interest lie between the regimes
of applicability of the Rayleigh approximation and geometric optics.
Noting that one convenient feature of the geometric optics approximation
is the easy visualisation of the fields (i.e., rays), it would be useful
to retain this feature in the intermediate regime.
For example, we can generate a full-wave equivalent to
\fref{fig:ashkin-ish-diagrams}, as shown in \fref{fig:deflection}.

There are a variety of computational methods that can achieve this.
Typically, they involve calculation of the fields over a finite computational
volume (which is what lends them so well to visualisation), and
the finite extent of this volume can make use of the
far-field limit impractical. Thus, the more general procedure described
above, using the MST, can be necessary.

\section{Finite difference time domain method}
\label{sec:fdtd}

The finite difference time domain method is a numerical method
for solving systems of partial differential equations that involves
storing discrete values for the fields at specific grid locations and
repeatedly applying the update equations to advance the fields though time.
For optical tweezers simulations, the partial differential equations
of interest are the time domain form of the source free Maxwell's equations,
\begin{equation}
\nabla\times E = -\frac{\partial B}{\partial t} \\
\end{equation}
\begin{equation}
\nabla\times H = \frac{\partial D}{\partial t} \\
\end{equation}
\begin{equation}
\nabla\cdot D = 0 \\
\end{equation}
\begin{equation}
\nabla\cdot B = 0
\end{equation}
where $D = \epsilon E$ and $B = \mu H$ are the real valued
electric and magnetic vector fields and $\epsilon, \mu$ are the
permittivity and permeability values for the field.
The discretisation step involves replacing the derivatives with
discrete differences, for example
\begin{equation}
\frac{\partial D}{\partial t} \quad\rightarrow\quad
    \epsilon \frac{E(t + \Delta t/2) - E(t - \Delta t/2)}{\Delta t} \\
\end{equation}
\begin{eqnarray}
\fl \left( \nabla\times H \right)_x
    = \frac{\partial H_z}{\partial y} - \frac{\partial H_y}{\partial z}
        \nonumber
    \\ \mkern-40mu \quad\rightarrow\quad
    \frac{H_z(y + \Delta y/2) - H_z(y - \Delta y/2)}{\Delta y}
    - \frac{H_y(z + \Delta z/2) - H_y(z - \Delta z/2)}{\Delta z}
\end{eqnarray}
where $\Delta t, \Delta y$ and $\Delta z$ are the time step size
and grid spacing along the $y$ and $z$ directions.

\begin{figure}[ht]
\centering
\includegraphics[width=\textwidth]{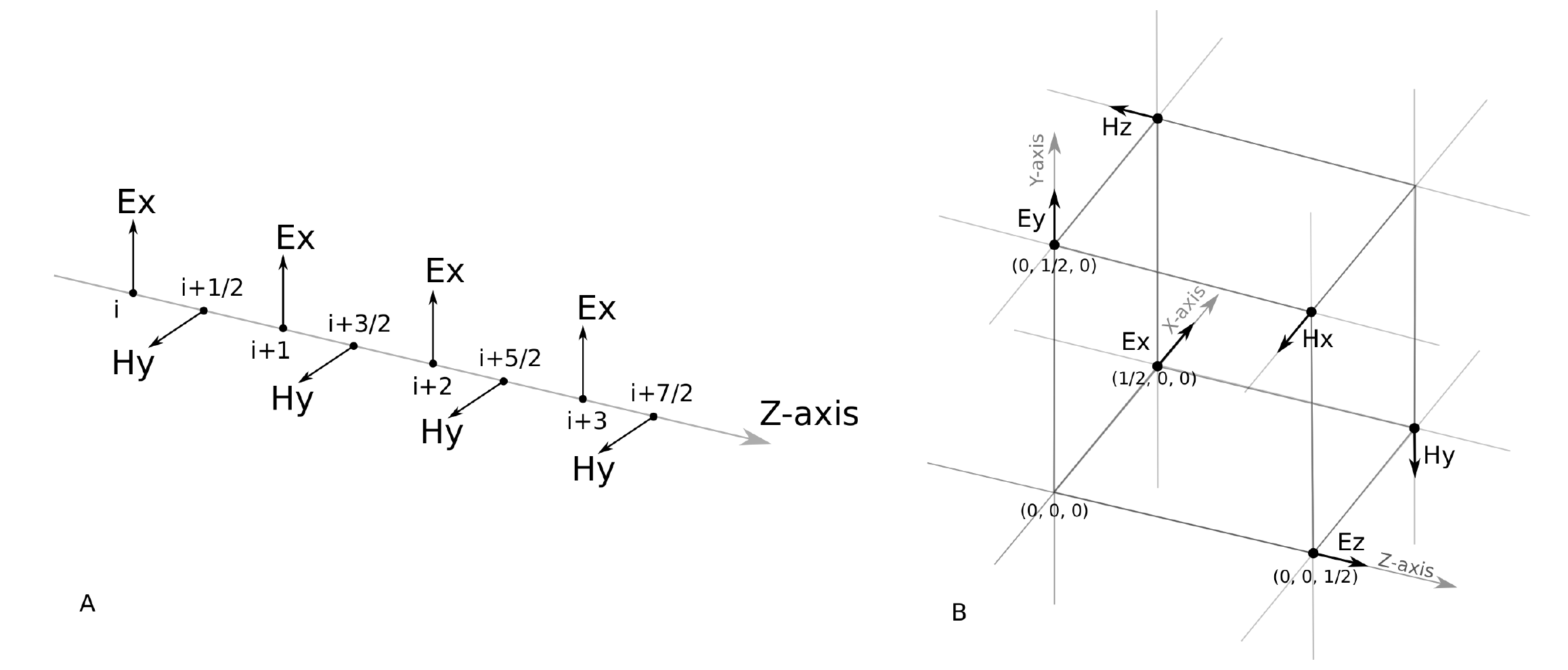}
\caption{(A) the offset grid used for 1-dimensional FDTD
    and (B) the Yee cell used for 3-dimensional FDTD.}
\label{fig:fdtd-offset-grid}
\end{figure}

For a 1-Dimensional transverse electromagnetic wave propagating
along the $z$-axis, with the
$E$ field oscillating along the $x$-axis, the corresponding equations
are
\begin{equation}
E^{n + 1}_{k + 1/2} = E^{n}_{k + 1/2}
    - \frac{\Delta t}{\epsilon}
      \frac{H_{k + 1}^{n + 1/2} - H_{k}^{n + 1/2}}{\Delta z},
\label{eq:isotropic-e-update}
\end{equation}
\begin{equation}
H^{n + 1/2}_{k} = H^{n - 1/2}_{k} 
    - \frac{\Delta t}{\mu}
      \frac{E_{k + 1/2}^{n} - E_{k - 1/2}^{n}}{\Delta z}.
\label{eq:isotropic-h-update}
\end{equation}
The subscript refers to the grid location and the superscript
refers to the time value.
The method achieves second order numerical accuracy by using
second order accurate central differences.
Rather than store the $E$ and $H$ fields at every grid location,
the fields are stored on a staggered grid, as shown in
\fref{fig:fdtd-offset-grid}.
For the three dimensional Cartesian case, the arrangement is
referred to as the Yee cell.
In addition to the grid being spatially staggered, the update equations
are also applied at alternate half-integer time steps to achieve
second order numerical accuracy for the evolution through time.
This approach is sometimes referred to as a leapfrog scheme/method.

Optical forces can be calculated from the $E$ and $H$ fields
using either the surface or volume integral in
\eref{eq:time-averaged-force} and \eref{eq:volume-integral-force}.
The FDTD method does not calculate all the components of
the $E$ and $H$ field at each location.
In order to evaluate the MST, the $E$ and $H$ field values surrounding
the particle are averaged to produce the necessary components for
the $E$ and $H$ fields at the location of interest.

The method requires the fields to be stored at every location throughout
the region of interest, such as in and around the particle.
At the edges of the simulation space the update equations for the
points along the edges depend on terms outside the simulation space.
This problem can be resolved in a number of ways: the values outside
the simulation space can simply be omitted or the derivatives at
the edge set to zero; periodic boundary conditions can be employed
where the values from the opposite side of the simulation space
are used for the missing values; or absorbing boundary conditions
can be used that will absorb any illumination incident on the boundary.

FDTD is of interest for two main reasons: the scalability with problem
size and its generality.
The method is also convenient if the fields are desired, since
the method requires storage of the field values at each grid location.
For a spherical particle the number of points and correspondingly the
amount of memory required to store the particle scales with the
volume $N^3$.
The total runtime for the simulation depends on the number of points
that need to be updated and the time it takes for a wave to propagate
across the simulation space which for a spherical particle is proportional
to the radius of the particle, so the time scales with $N^4$.

The other advantage of FDTD is its generality.
The above derivation for the 1-dimensional case assumed an isotropic
permittivity, however the method can be extended for anisotropic,
conductive and even non-linear materials \cite{benito08, nefdtd11}.
The shapes of particles the method can simulate is only dependent on
the size and required resolution of the particle features, which in
turn depends on the scalability of the method.
Multiple types of illumination can be using including plane wave beams
and tightly focussed Gaussian beams from pulsed, broadband or
continuous sources.
For optical tweezers simulations, the main property the method
can calculate is optical forces.
The method can also be used to calculate various other
properties of potential interest to optical tweezers research
including torques, internal stresses/strains and absorption/heating.

\subsection{Types of illumination}

Common types of illumination used for optical tweezers
include tightly focussed Gaussian beams, Laguerre--Gaussian beams
and evanescent fields at the boundaries of high-low refractive
index materials.
Situations such as \fref{fig:apparatus}-A and \ref{fig:apparatus}-D
can be modelled by describing the incident illumination at one side
of the simulation and running the simulation until the fields reach
steady state for continuous wave illumination or until the illumination
leaves the simulation region for pulsed beams.
In some situations it may be necessary to also model parts of the
apparatus in order to properly describe the incident illumination or the
reflections between the particle and the apparatus.
Evanescent fields (\fref{fig:apparatus}-B/C)
can be modelled by introducing the illumination,
such as a plane wave or Gaussian beam, onto the interface between
high and low refractive index materials at a suitable angle to cause
total internal reflection.

A variety of methods exist to introduce illumination sources into
FDTD simulations, they fall broadly into two categories: hard sources,
which scatter/reflect any fields incident on the source location;
and soft sources which allow other fields to pass through the source location.
One type of soft source is the total field scattered field (TFSF)
which is so named because it separates the simulation space into
a region that contains only the scattered fields and a region that
contains the total fields (scattered + incident) \cite{nefdtd11}.
In this paper we have used TFSF for introducing sources into our simulations.
TFSF allows the incident beam to be introduced on a surface
surrounding the particle as long as the values for the $E/H$ fields
are known everywhere on the surface.
As such, the problem of describing the incident illumination is
reduced to a problem of finding an expression for the $E/H$ fields
on the surface.

If the $E/H$ field amplitudes are known/computed prior,
perhaps as a result of a previous FDTD simulation and stored in a file,
the process of introducing the source involves reading in values
from the file, interpolating the time/grid coordinates to match
the FDTD simulation and adding the fields to the corresponding
grid locations.
However, many types of beams can be described more simply by their
complex field amplitudes or as a decomposition of amplitudes in some basis.
For example, a linearly polarized plane wave beam propagating
along the $\vec{z}$ direction can be described by
the time harmonic complex amplitudes
\begin{equation}
\mathbf{E} = e^{ikz} \vec{x}, \qquad
\mathbf{H} = \sqrt{\frac{\epsilon}{\mu}} e^{ikz} \vec{y}.
\label{eq:plane-wave-beam}
\end{equation}
FDTD requires the real field amplitudes, which
can be easily calculated from the complex amplitudes
\begin{equation}
E(t) = \Re\left[e^{i\omega t} \mathbf{E} \right], \qquad
H(t) = \Re\left[e^{i\omega t} \mathbf{H} \right].
\label{eq:complex-to-real-fields}
\end{equation}
These field values can then be added to the
appropriate grid locations at each time step.

As with plane wave beams, Gaussian beams can also be described
by a set of complex field amplitudes.
A common method involves describing the Gaussian beam in some
basis, such as in terms of the vector spherical wave functions,
where the $\mathbf{E}$ and $\mathbf{H}$ field amplitudes
are given by
\begin{equation}
\mathbf{E} = \sum_{n = 1}^\infty \sum_{m = -n}^n
  a_{nm} \mathrm{Rg} M_{nm}(kr, \theta, \phi)
+ b_{nm} \mathrm{Rg} N_{nm}(kr, \theta, \phi)
\label{eq:vswf-e}
\end{equation}
\begin{equation}
\mathbf{H} = -j \sqrt{\frac{\epsilon}{\mu}} \sum_{n = 1}^\infty \sum_{m = -n}^n
  b_{nm} \mathrm{Rg} M_{nm}(kr, \theta, \phi)
+ a_{nm} \mathrm{Rg} N_{nm}(kr, \theta, \phi)
\label{eq:vswf-h}
\end{equation}
where $(kr, \theta, \phi)$ are the spherical coordinates centred at the
beam focus, and $a_{nm}, b_{nm}$ are coefficients describing the
beam in terms of the regular vector spherical wave functions, VSWFs,
$\mathrm{Rg} M_{nm}, \mathrm{Rg} N_{nm}$ \cite{benito08}.
A particular beam is then described by a particular set of beam coefficients.
Coefficients for Gaussian beams can be calculated for weakly focussed
beams using the paraxial approximation and for tightly focussed beams
by adding higher order corrections to the paraxial approximation,
as is done for the fifth order Davis beam \cite{barton89, davis79}
used by \cite{benito08}.
Another approach is to use an over-determined point matching method
to determine the beam coefficients \cite{nieminen03}.
Matlab codes are available as part of the optical tweezers toolbox
for computing the beam coefficients for tightly focussed beams including the
Laguerre-Gaussian modes with angular momentum \cite{nieminen14}.
Instead of using a basis of vector spherical wave functions
other approaches describe the incident illumination with a plane wave basis.
Reference \cite{capoglu13} model tightly focussed Hermite-Gaussian beams
in FDTD with TFSF using a plane wave basis.

\begin{figure}[ht]
\centering
\includegraphics[width=0.24\textwidth,height=0.24\textwidth]{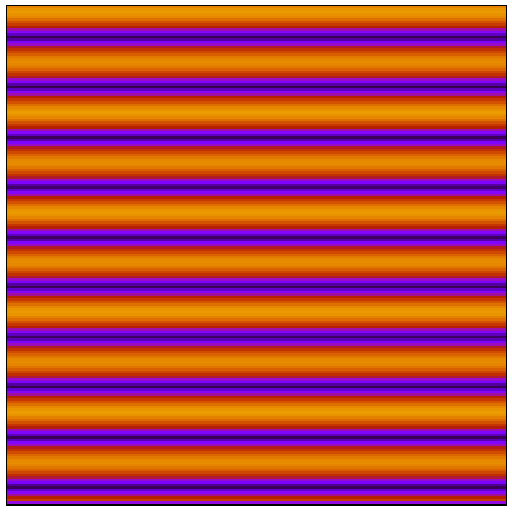}\hfill
\includegraphics[width=0.24\textwidth,height=0.24\textwidth]{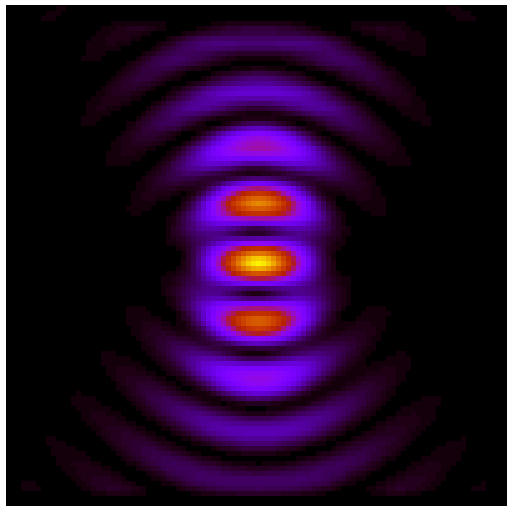}\hfill
\includegraphics[width=0.24\textwidth,height=0.24\textwidth]{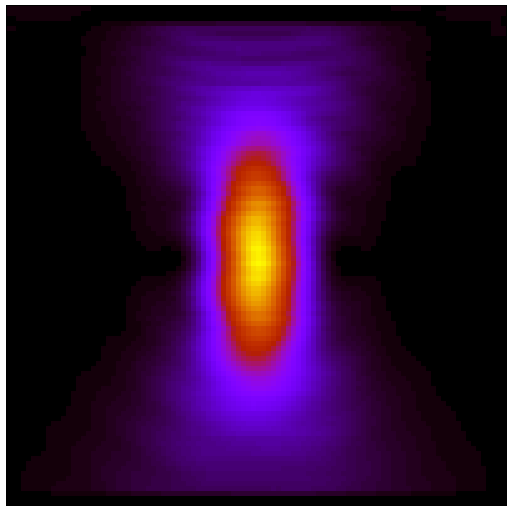}\hfill
\includegraphics[width=0.24\textwidth,height=0.24\textwidth]{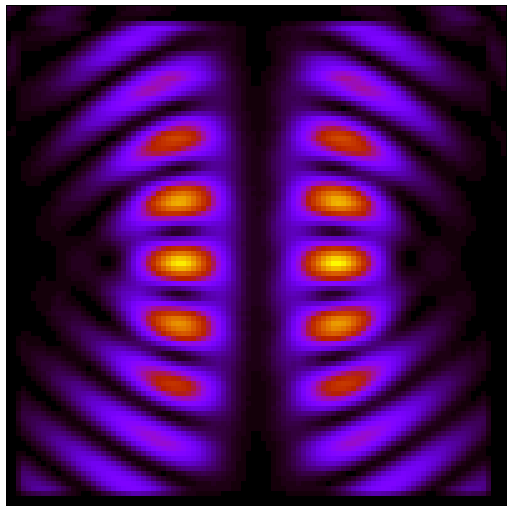}\\
\includegraphics[width=0.24\textwidth,height=0.24\textwidth]{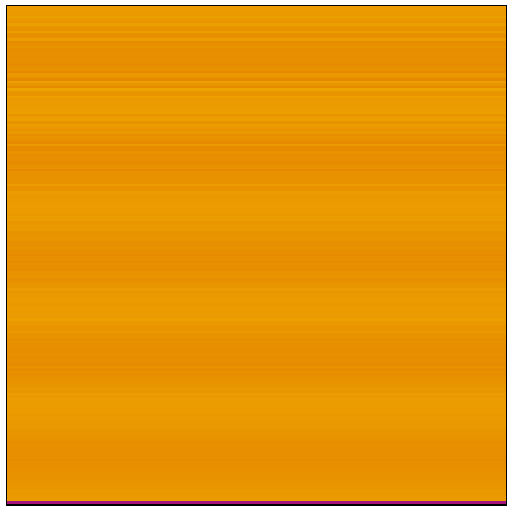}\hfill
\includegraphics[width=0.24\textwidth,height=0.24\textwidth]{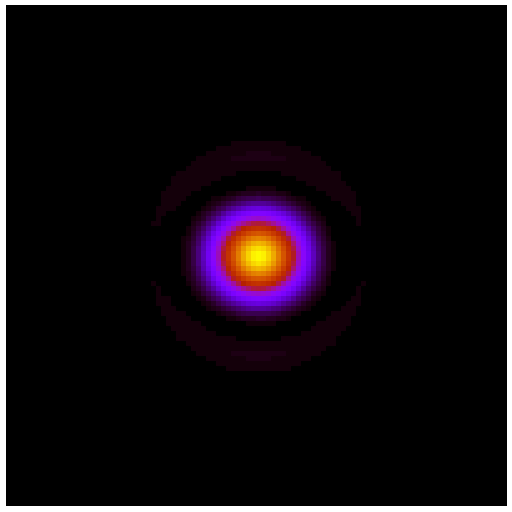}\hfill
\includegraphics[width=0.24\textwidth,height=0.24\textwidth]{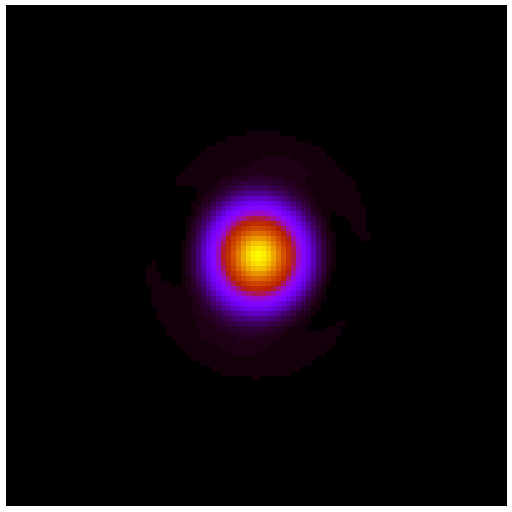}\hfill
\includegraphics[width=0.24\textwidth,height=0.24\textwidth]{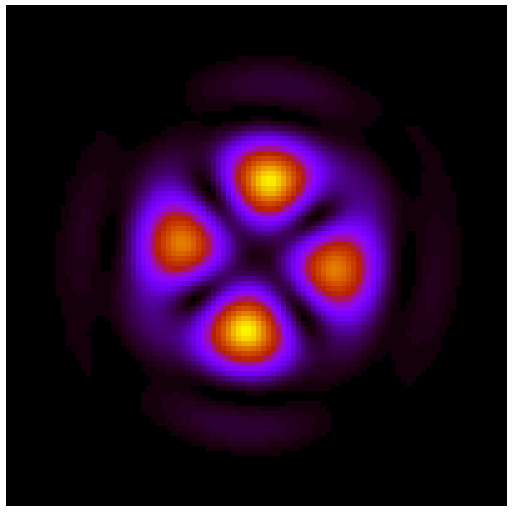}
\caption{Different types of illumination.  Top (left-to-right):
    Linearly polarized plane wave, linearly polarized focussed Gaussian beam,
    circularly polarized focussed Gaussian beam, linearly polarized LG02 beam.
    Bottom: circularly polarized plane wave (left), focal plane
    images of the beam spot for the above beam types.}
\label{fig:illumination}
\end{figure}

\Fref{fig:illumination} shows plane wave beams, focussed Gaussian
beams and Laguerre-Gaussian with different polarisations.
The full fields can be visualised in a number of ways including
plotting the power density, $E/H$ field components or the $E/H$
field intensity.
In this paper we plot the instantaneous $E$ field intensity $|E|^2$.
For linearly polarized beams this allows the beam wave fronts to
be visualised.
For circularly polarized beams the field pattern resembles the
time averaged field intensity $\langle|E|^2\rangle$.
When a scattering particle is introduced the reflected
light causes interference fringes that can be seen in
the $|E|^2$ visualisation.

\section{Simulations}
\label{sec:simulations}

In this section we present the fields and optical forces calculated
using the FDTD method for different optical trapping scenarios.
The section contains six groups of simulations.
Four scenarios involve spherical particles with different sizes
and refractive indexes in circular and linearly
polarized tightly focussed Gaussian beams.
The other two scenarios involve evanescent fields above a
high-lower refractive index interface.
Calculated forces are normalized by the beam power and speed
in medium surrounding the particle to give the dimensionless
force efficiencies.
The dimensionless force efficiency describes the force in units
of $n\hbar k_0 = \hbar k$ per photon.

\begin{figure}[htp]
\centering
\hfill\includegraphics[width=0.31\textwidth]{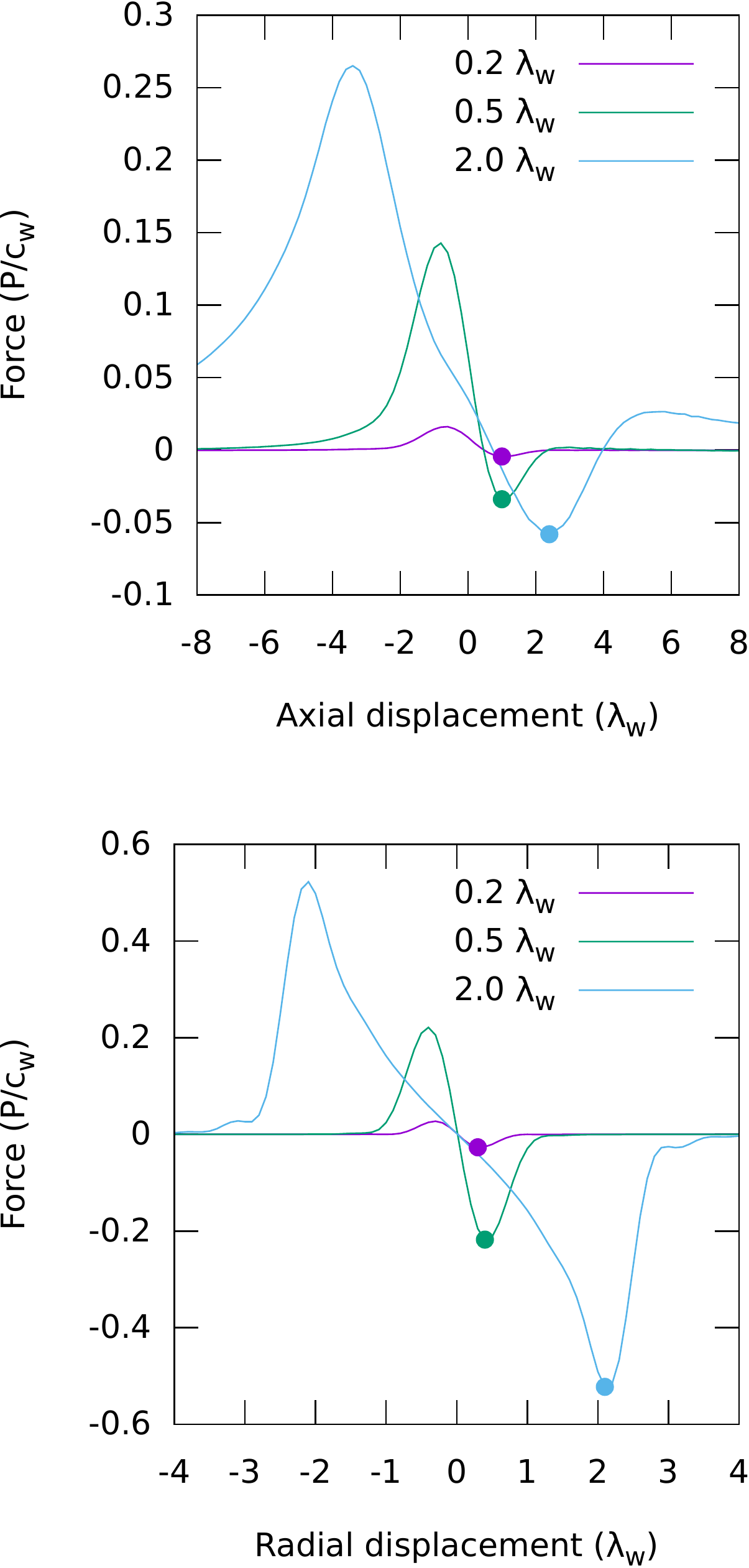}\hfill
\begin{minipage}[b]{0.2\textwidth}
\includegraphics[width=\textwidth]{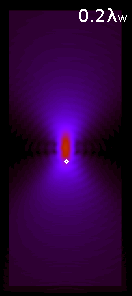}\\
\includegraphics[width=\textwidth]{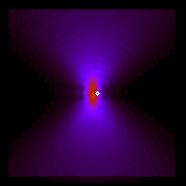}
\end{minipage}
\begin{minipage}[b]{0.2\textwidth}
\includegraphics[width=\textwidth]{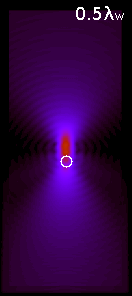}
\includegraphics[width=\textwidth]{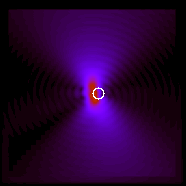}
\end{minipage}
\begin{minipage}[b]{0.2\textwidth}
\includegraphics[width=\textwidth]{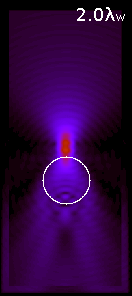}\\
\includegraphics[width=\textwidth]{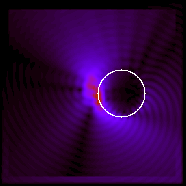}
\end{minipage}
\caption{Different sized polystyrene spheres in water.}
\label{fig:simulation-sizes}
\end{figure}

\Fref{fig:simulation-sizes} shows different sized polystyrene
spheres suspended in water illuminated by a circularly polarized
focussed Gaussian beam.
The numerical aperture (NA) of the beam is 1.02.
The illumination wavelength is 800nm in water ($\lambda_w$),
the refractive index of water and polystyrene are 1.33 and 1.59 respectively.
The simulation of the $0.2\lambda_w$ particle shows the beam is almost
completely unaffected by the presence of the particle.
The force the particle experiences is proportional to the gradient of
the field amplitude. At the centre of the beam, the gradient is almost
zero, and further out the gradient increases in magnitude before falling off
quickly as the beam spreads out.
For the larger particles the deflection and scattering of the beam
is more noticeable; for the $2\lambda_w$ axial figure the beam is clearly
more focussed.

\begin{figure}[htp]
\centering
\begin{minipage}[t]{0.31\textwidth}
\raisebox{-0.53\height}{
\includegraphics[width=0.9\textwidth]{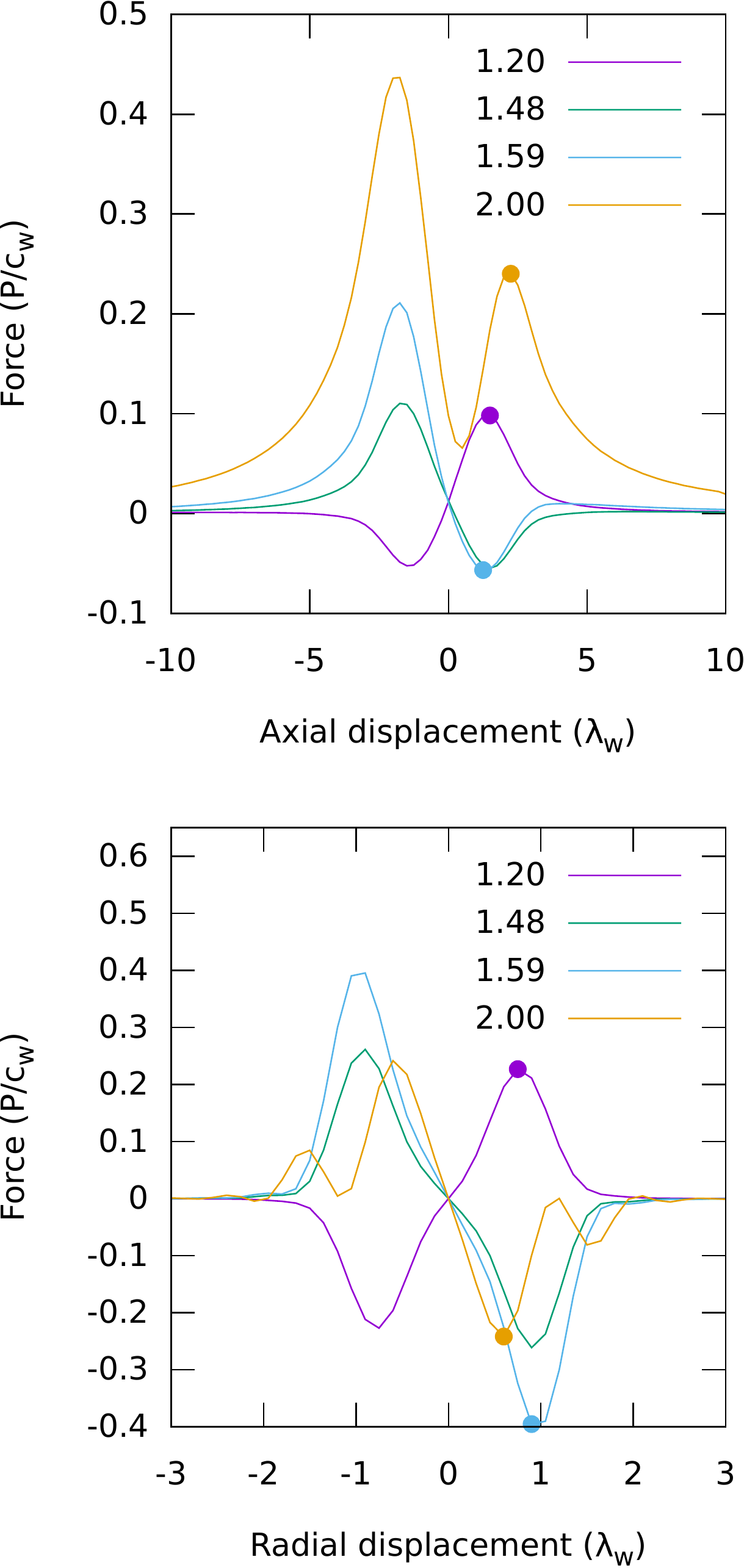}
}
\end{minipage}
\begin{minipage}[t]{0.22\textwidth}
\includegraphics[width=\textwidth]{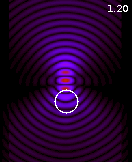}
\includegraphics[width=\textwidth]{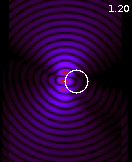}
\end{minipage}
\begin{minipage}[t]{0.22\textwidth}
\includegraphics[width=\textwidth]{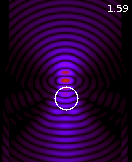}
\includegraphics[width=\textwidth]{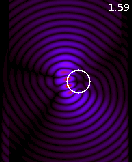}
\end{minipage}
\begin{minipage}[t]{0.22\textwidth}
\includegraphics[width=\textwidth]{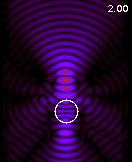}
\includegraphics[width=\textwidth]{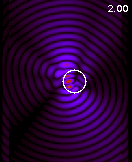}
\end{minipage}
\caption{Spheres with different refractive indices suspended in water.}
\label{fig:simulation-index}
\end{figure}

\Fref{fig:simulation-index} shows the effect of different refractive
indices on the optical force and scattering of the beam by the
dielectric sphere.
The spheres are illuminated by a linearly polarized Gaussian beam
(NA = 1.02), the $E$-field is polarized parallel to the radial
displacement direction.
The high index (n = 2.0) and very low index (n = 1.2) cases
have force-displacement curves that do not correspond to optical trapping.
The very low index sphere has a negative force-displacement curve,
in the radially displaced field simulation the particle appears to
repel the beam, in the axial simulation the beam becomes less collimated.
The light that passes through the high index sphere is very collimated,
but much more of the light is reflected off
the first surface and scattered in other directions.

\begin{figure}[htp]
\centering
\hfill\raisebox{-0.51\height}{\includegraphics[width=0.28\textwidth]{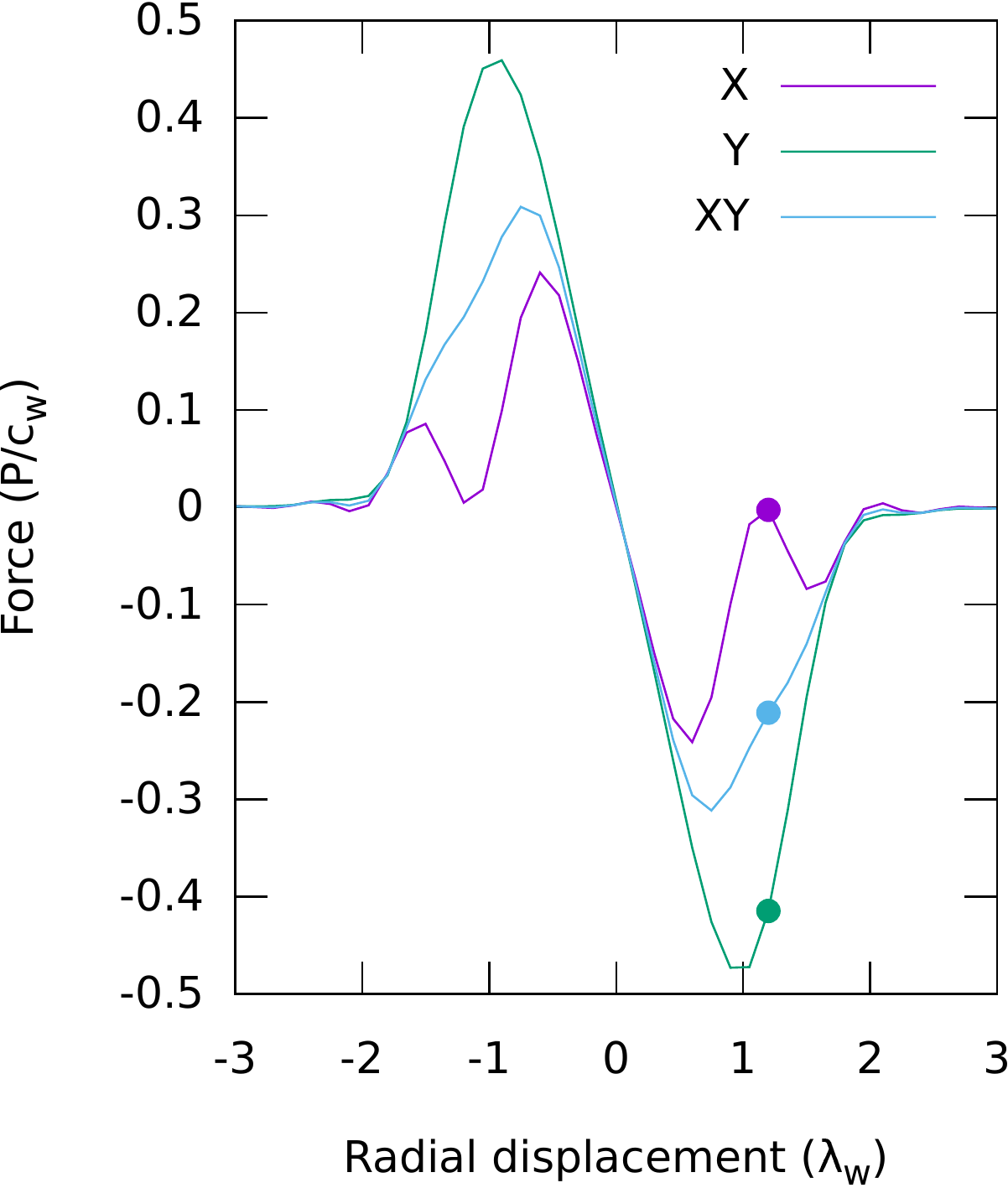}}\hfill
\hfill\raisebox{-0.45\height}{\includegraphics[width=0.22\textwidth]{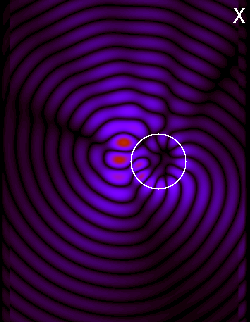}}\hfill
\hfill\raisebox{-0.45\height}{\includegraphics[width=0.22\textwidth]{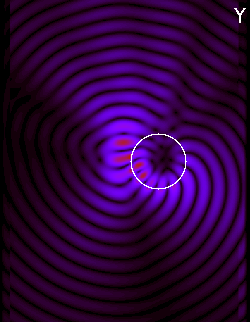}}\hfill
\hfill\raisebox{-0.45\height}{\includegraphics[width=0.22\textwidth]{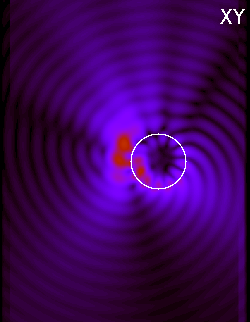}}\hfill
\caption{2$\lambda_w$ high index (n = 2.0) sphere in water illuminated by
    linear and circularly polarised Guassian beams.}
\label{fig:simulation-polarisation}
\end{figure}

The effect of beam polarisation on the trapping of particles can be
important, especially when the surface of the particle is near the
edge of the beam.
\Fref{fig:simulation-polarisation} shows the effect of polarisation
on the radial force on a high index sphere in water.
When the light is incident near normal to the surface, the transmittance
between the water and sphere will be almost equal for all
polarisation angles.
As the incident angle approaches Brewster's angle, the reflectivity
between light polarised perpendicular and parallel to the incident
plane becomes noticeably different.
In the figure, when the beam is polarised along the \texttt{Y} axis,
much more of the light appears to pass into the sphere than in
the \texttt{X} case where most of the beam appears to be deflected
around the sphere.
The fields in the circularly polarized case, with the same total beam power,
resemble the sum of the two linear cases; this is also true for
the time averaged force.

\begin{figure}[htp]
\centering
\begin{minipage}[t]{0.31\textwidth}
\raisebox{-0.53\height}{
\includegraphics[width=0.9\textwidth]{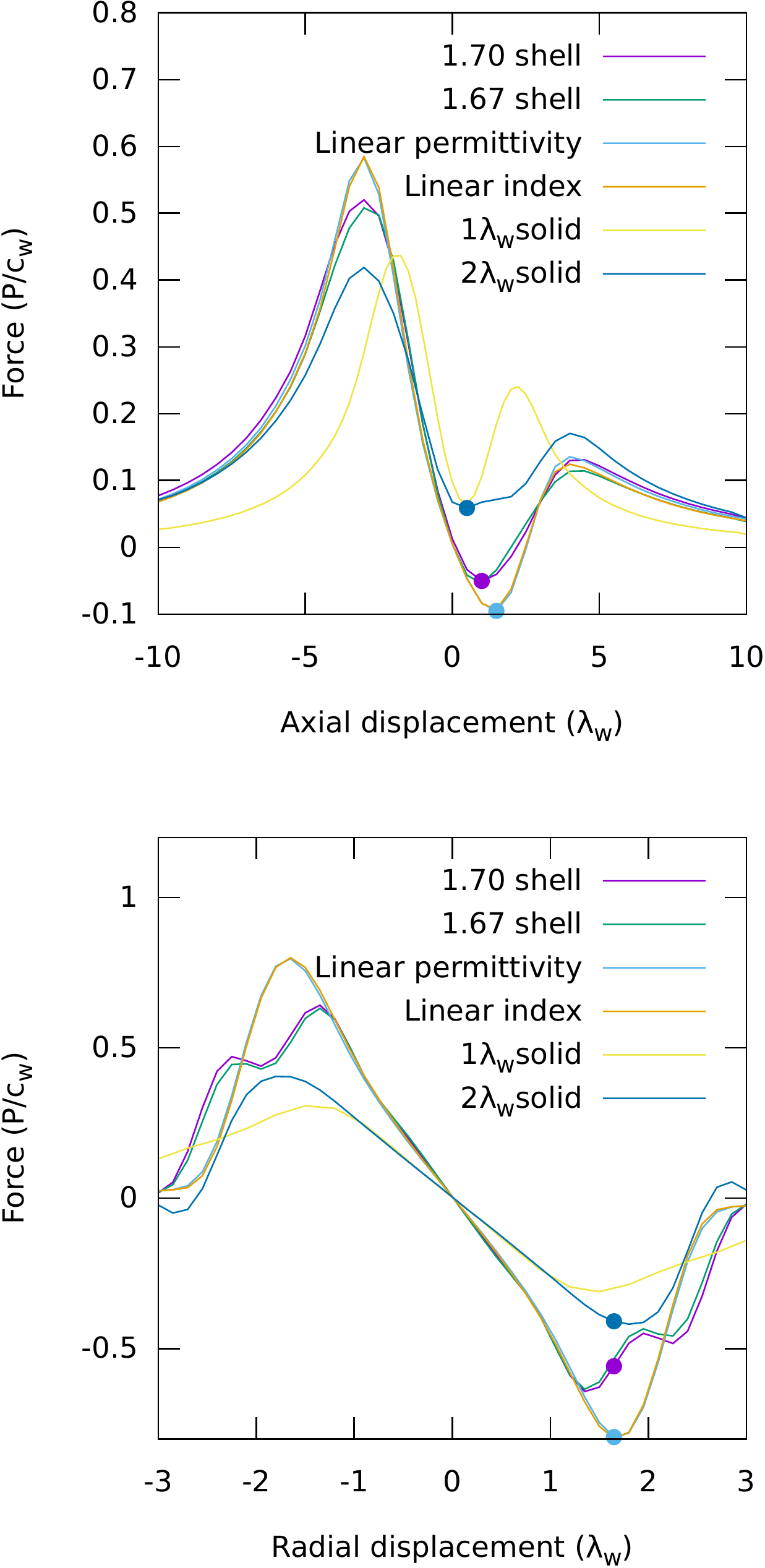}
}
\end{minipage}
\begin{minipage}[t]{0.22\textwidth}
\includegraphics[width=\textwidth]{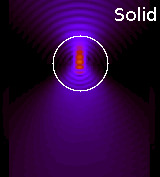}
\includegraphics[width=\textwidth]{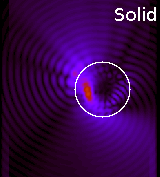}
\end{minipage}
\begin{minipage}[t]{0.22\textwidth}
\includegraphics[width=\textwidth]{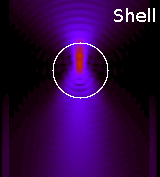}
\includegraphics[width=\textwidth]{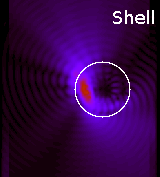}
\end{minipage}
\begin{minipage}[t]{0.22\textwidth}
\includegraphics[width=\textwidth]{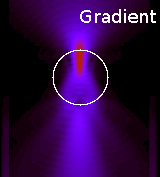}
\includegraphics[width=\textwidth]{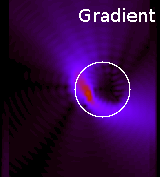}
\end{minipage}
\caption{Spherical $1\lambda_w$ high index (n = 2.0) sphere coated with
    different anti-reflective coatings.  The \texttt{Solid} case
    shows a $2\lambda_w$ high index sphere for comparison.}
\label{fig:simulation-ar-coated}
\end{figure}

The previous figures showed that it is sometimes difficult to
trap spheres when there is a high refractive index contrast between
the trapping medium and the particle.
If it is not possible to change the refractive index of the background
medium, one alternative is to coat the particle with an
anti-reflective coating.
This technique has been demonstrated experimentally for polystyrene
spheres coated with silica \cite{bormuth08}.
\Fref{fig:simulation-ar-coated} shows 4 different types of engineered
particles with anti-reflective (AR)
coatings surrounding a $1\lambda_w$ radius high index (n = 2.0) sphere.
The AR coatings form a $1\lambda_w$ shell around the sphere.
For comparison a $2\lambda_w$ high index sphere is also shown.

We thought it would be interesting to try multiple types of AR
coatings: two with a single layer of material and two with a graded material.
The refractive index of the single layer AR coatings were chosen to
be half way between the refractive index of the two materials and
half way between the permittivity values of the two materials.
The grading was chosen to be linear with respect to the refractive index
for one case and linear with respect to the permittivity for the other.
The graded AR coatings performed best.
For this particular scenario we did not notice any significant
difference between the two choices of outer refractive index
and the two choices of gradings.
This result agrees with the observation by \cite{yinghu08} that for
single layer coated spheres, improvements to the trapping
efficiency are relatively insensitive to refractive index and
thickness of the surrounding layer.

\begin{figure}[htp]
\centering
\hfill
\begin{minipage}[t]{0.31\textwidth}
\raisebox{-0.70\height}{
\includegraphics[width=\textwidth]{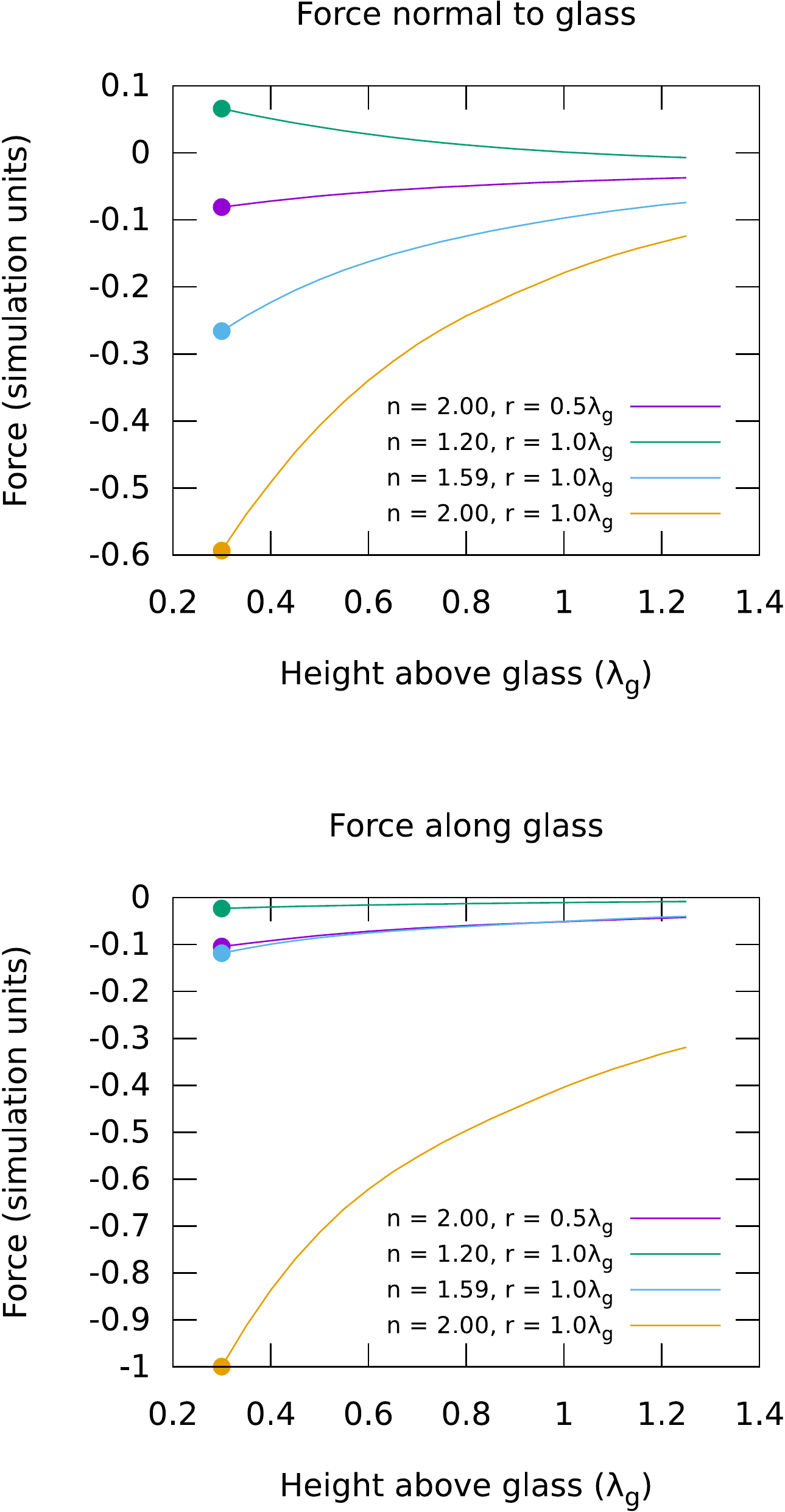}
}
\end{minipage}
\hfill
\begin{minipage}[t]{0.6\textwidth}
\includegraphics[width=\textwidth]{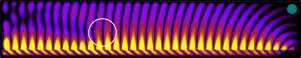}
\includegraphics[width=\textwidth]{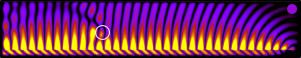}
\includegraphics[width=\textwidth]{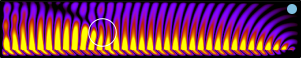}
\includegraphics[width=\textwidth]{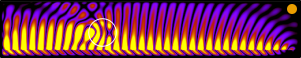}
\end{minipage}
\hfill
\caption{Particles in an evanescent transport beam.}
\label{fig:simulation-ev-transport}
\end{figure}

\begin{figure}[htp]
\centering
\hfill
\begin{minipage}[t]{0.31\textwidth}
\raisebox{-0.65\height}{
\includegraphics[width=\textwidth]{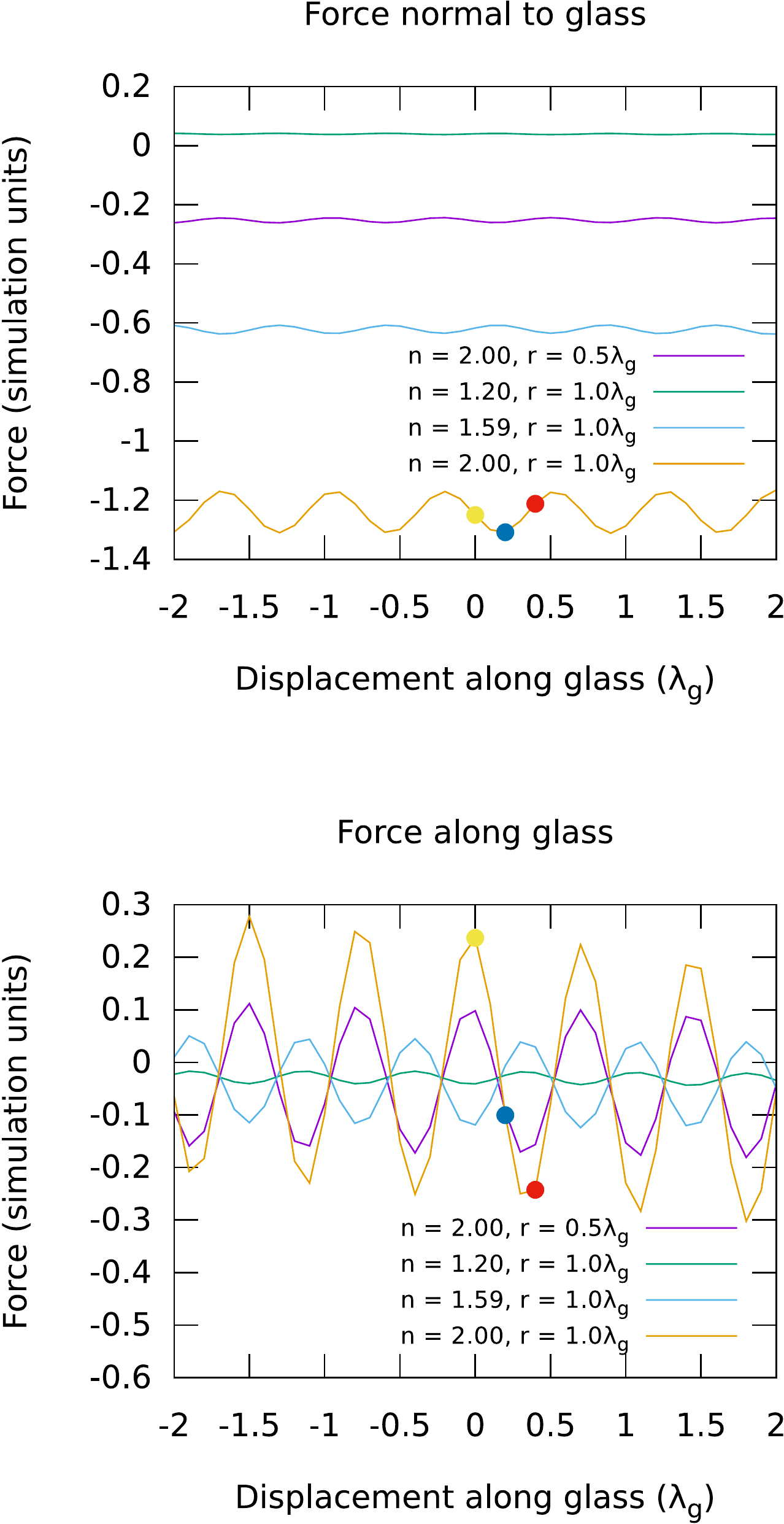}
}
\end{minipage}
\hfill
\begin{minipage}[t]{0.6\textwidth}
\includegraphics[width=\textwidth]{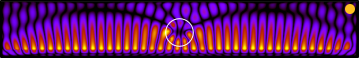}\\[1em]
\includegraphics[width=\textwidth]{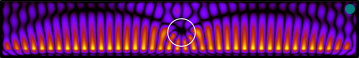}\\[1em]
\includegraphics[width=\textwidth]{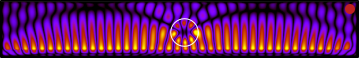}
\end{minipage}
\hfill
\caption{Particles in an evanescent standing wave beam.}
\label{fig:simulation-ev-standing}
\end{figure}

The two evanescent field simulations are shown in
\fref{fig:simulation-ev-transport} and  \fref{fig:simulation-ev-standing}.
These simulations used an artificial beam located beneath the high-low
refractive index interface ($n_{high} = 2.0, n_{low} = 1.33$).
The illuminating beam was a plane wave angled at 43.2 degrees from the
surface normal.
The calculated forces for the evanescent simulations are all
normalized to the maximum force value in \fref{fig:simulation-ev-transport}.

\Fref{fig:simulation-ev-transport} involves a single circularly
polarized beam.
The beam is propagating from right to left.
On the far right there are additional non-evanescent fields above the
surface, these are an artifact of the beam, similar but smaller artifacts
are also present on the far left of the simulation.
The evanescent field appears to be coupled into the high index
spheres and subsequently be radiated outwards.
As a result, the spheres experience a force towards the surface and
along the direction of the field.
The low index sphere appears to repel the field back down towards
the surface, the sphere is repelled from the surface but still
pushed along by the field.

\Fref{fig:simulation-ev-standing} involves two linearly polarized
illumination sources propagating from opposite ends of the simulation.
Similarly to \fref{fig:simulation-ev-transport} the simulation contains
artifacts towards the far edges of the simulation due to the type of
beam and simulation constraints.
Unlike the single beam evanescent field, with multiple beams it is
possible to trap and hold a particle in place.
Depending on the size and refractive index of the sphere, stable trapping
may correspond to the dark spot between evanescent field fringes or
the light spot at an evanescent field fringe.

\section{Conclusion}

In this paper we have discussed optical tweezers, the different
regimes for modelling optical tweezers problems and simulations
of different scenarios calculated using FDTD.
The full field visualisations show the qualitative behaviour of
the fields around a scattering particle, providing a similar qualitative
insight into the trapping behaviour as the Rayleigh
approximation and geometric optics methods with the advantage of
providing accurate results in the intermediate regime.
Unlike geometric optics, the full field visualisation approach is
also able to provide insight into the behaviour of particles in
evanescent fields, visually showing the coupling of the field
out of the high index material.

FDTD is a useful tool for visualising the fields in and around
arbitrary scatterers, the method is very general and can be used
for modelling various types of various materials.
In this paper we limited our study to particles with spherical
symmetry.
Particles without rotational symmetry may also experience optical
torques from transfer of spin or orbital angular momentum from the
beam to the particle.
Calculation of optical torques in the near-field using FDTD is
difficult. Taking the moment of the force per unit volume
using \eref{eq:volume-integral-force} seems like a good starting
points, but this approach seems to omit the spin angular momentum.
Alternatively, \cite{benito08} use another stress tensor approach
that appears to include the spin in certain circumstances while omitting
the orbital angular momentum.
It is unclear if a combination of these two approaches can be
used to give accurate results.
The higher dimensionality also leads to increased difficulty in
finding/optimising trapping positions and conditions.
For these higher dimensional search spaces an optimisation technique
such as simulated annealing might be useful.

The simulations presented in this paper used our new FDTD implementation.
Our FDTD implementation is written in C++ utilising newer features
of the language including template meta-programming.
It is our plan to release the implementation as open source
software in the near future.

\ack
Parts of this work were undertaken by I C D Lenton as part of
the Bachelor of Science (Honours) program at The University of Queensland.
This research was supported under Australian Research Council's
Discovery Projects funding scheme (project number DP140100753).

\section*{References}


\providecommand{\newblock}{}

\end{document}